\documentclass[apj,twocolumn]{emulateapj}

\usepackage{amsmath}

\slugcomment{Accepted for publication in ApJ}

\shortauthors{G. Parmentier}
\shorttitle{The variations of the \sfe per \fft }

\newcommand\eff{\epsilon_{\rm ff}}
\newcommand\effint{\epsilon_{\rm ff, int}}
\newcommand\effmeas{\epsilon_{\rm ff, meas}}

\newcommand\tff{\tau_{\rm ff}}

\newcommand\Ms{M_{\odot}}

\newcommand\Mspp{M_{\odot} \cdot pc^{-2}}
\newcommand\Msppp{M_{\odot} \cdot pc^{-3}}

\newcommand\cc{cm^{-3}}

\newcommand\fft{free-fall time }
\newcommand\sfe{star formation efficiency }
\newcommand\sfr{star formation rate }
\newcommand\stf{star formation }
\newcommand\sfing {star-forming }

\begin{document}



\title{The density gradient inside molecular-gas clumps \\ as a booster of their star formation activity}


\author{G.~Parmentier\altaffilmark{1}}


\altaffiltext{1}{Astronomisches Rechen-Institut, Zentrum f\"ur Astronomie der Universit\"at Heidelberg, M\"onchhofstr. 12-14, D-69120 Heidelberg, Germany}


\begin{abstract}
Star-forming regions presenting a density gradient experience a higher star formation rate than if they were of uniform density.  We refer to the ratio between the star formation rate of a spherical centrally-concentrated gas clump and the star formation rate that this clump would experience if it were of uniform density as the magnification factor $\zeta$.  We map $\zeta$ as a function of clump mass, radius, initial volume density profile and star formation time-span.  For clumps with a steep density profile (i.e. power-law slope ranging from $-3$ to $-4$, as observed in some high-density regions of Galactic molecular clouds), we find the star formation rate to be at least an order of magnitude higher than its top-hat equivalent.  This implies that such clumps experience faster and more efficient star formation than expected based on their mean free-fall time.  This also implies that measurements of the star formation efficiency per free-fall time of clumps based on their global properties, namely, mass, mean volume density and star formation rate, present wide fluctuations.  These reflect the diversity in the density profile of \sfing clumps, not necessarily variations in the physics of star formation.  Steep density profiles inside star-cluster progenitors may be instrumental in the formation of multiple stellar populations, such as those routinely observed in old globular clusters.   \\
\end{abstract}



\section{Introduction} \label{sec:intro}
The process of star formation is the engine of galaxy evolution.  Galaxies form a significant fraction of their stars inside  clumps of dense molecular gas, themselves embedded in more diffuse giant molecular clouds, with respective spatial scales of 1\,parsec and a few tens of parsecs.  
Distant galaxies, molecular clumps of the Galactic disk and molecular clouds of the Solar neighbourhood present a linear correlation between their dense-gas mass and \stf rate \citep{gs04,wu10,lad10}. 
It thus seems that the activity of a \sfing region is controlled by its  dense-gas content, usually defined as gas with a number density higher than $\sim 10^4\,\cc$, equivalent to a mass density of $700\,\Msppp$. 
Yet, this correlation comes with a scatter.  Is this scatter driven by observational uncertainties only?  Or does it conceal secondary, yet to be discovered, properties?  Consider for instance two \sfing regions with different \stf rates.  Does the more active \sfing region necessarily contain a larger dense-gas mass?  Or could its higher \sfr result from its dense-gas clumps being more efficient at forming stars than those of the less active region, without modification of the overall dense-gas content?  

While the interplay between the \sfr of molecular clouds and the spatial distribution of their gas has been well-studied, both observationally  \citep[see e.g. ][]{lad13,kai14} and theoretically \citep{cho11,elm11}, the interplay between star formation rate and gas spatial distribution of small-scale roughly-spherical molecular clumps has received less attention.  In the disk of our Galaxy, the density profile inside molecular clumps obeys a power-law $\rho(r) \propto r^{-p}$, with $\rho(r)$ the gas volume density at a distance $r$ from the clump center and $p$ observed to range from $1.2$ to $2.5$ \citep{mue02}. 
The \sfr of clumps with $p\simeq(1.5,2.0)$ is predicted to be 1.6-to-2 times higher than the \sfr of clumps containing the same gas mass inside the same radius, but with a top-hat profile  ($p=0$). 
Assuming a pure power-law of slope $-1.5$ for the volume density profile of clumps, \citet{tan06} predict a \sfr 1.6 times higher than if they were of uniform density \citep[see eq.~2 in][]{tan06}.  For a clump with a density profile of slope $-2$ combined to a small ($\simeq 10^{-2}$ the clump radius) central core, \citet{par14p} finds that the \sfe achieved within one free-fall time amounts to $1.6\eff$, rather than $\eff$, with $\eff$ the \sfe per free-fall time.  Additionally, Fig.~9 in \citet{par13} shows that the \sfe of a clump increases faster with time when the clump density profile is steeper.  
Numerical simulations confirm these analytical and semi-analytical insights.  \citet{gir11a} performed  hydrodynamical simulations of molecular clumps with an initial gas mass of 100\,$\Ms$, a radius of 0.1\,pc, various initial gas density profiles and turbulence seeds.  Their fig.~3 shows that, once \stf has started, the mass in sink particles grows faster  for an initial gas density profile of slope $-2$ than for a top-hat density profile.  Specifically, the rate at which the gas mass is accreted onto the sink particles is twice as high for the power-law density profile with a slope of $-2$ as it is for its top-hat equivalent.  
In other words, more centrally-concentrated clumps show a higher \sfr as a result of the higher gas volume densities and shorter free-fall times characterizing their inner regions.  Finally, similar conclusions were reached by \citet{elm11} based on the gas density-PDF.  Modelling the density PDF of a spherically-symmetric cloud as the convolution of the log-normal PDF from turbulence for cloud local regions (where  the gas volume density does not vary significantly) with the overall cloud density profile, he too concludes that the \sfr of centrally-concentrated clouds is higher than that of clouds of uniform density.  All studies above have considered gas density profiles no steeper than $\rho(r) \propto r^{-2}$.

But recent observations have reported much steeper density profiles in dense-gas clumps of the Galactic molecular clouds MonR2 and NGC6334, with $p\simeq(3,4)$ \citep{schn15}.  These unlock a different regime of \stf activity: when $p>2$, not only do clump inner regions present a faster pace of star formation, they also contain most of the clump mass.  This raises a tantalizing question: by how much do so steep density profiles increase the clump \sfr compared to uniform-density clumps with identical mass and radius?  A comprehensive mapping of this effect was still missing, as well as a method to quantify it once star formation has modified the initial gas distribution.  

Not only does a gas density gradient inside clumps impact their \stf rate, it also affects the \sfe per \fft that we measure for such clumps.  The star formation efficiency per free-fall time, $\eff$, is a dimensionless key parameter related to the \sfr of galaxies, clouds and clumps.  It quantifies the gas mass fraction that a gas reservoir turns into stars every free-fall time $\langle \tff \rangle$, with the free-fall time $\langle \tff \rangle$ calculated at the gas mean density.
A proxy to the \sfr of a galaxy, cloud, or clump is thus, with $m_{gas}$ its \sfing gas mass: 
\begin{equation}
SFR = \frac{\eff m_{gas}}{\langle \tff \rangle}
\label{eq:intro}
\end{equation}
\citep[see e.g.][]{kru05,kt07,eva09,lad10,mur11,kru12,vut16,och17}.

According to Eq.~\ref{eq:intro}, to increase the \sfr of a clump, one can increase either its mass or its volume density or its \sfe per free-fall time.  However, as was shown by \citet{tan06}, \citet{gir11a}, \citet{elm11} and \citet{par14p}, one can also steepen its volume density profile.  That should be taken into account when comparing \sfing regions based on their \sfe per free-fall time.  Whether $\eff$ is constant or not remains debated.  \citet{kt07} argue that its value is about 0.01, constant in the Galactic disk across a wide range of volume densities, from low-density giant molecular clouds to high-density CS clumps (see their fig.~5).  In contrast, \citet{och17} argue that significant variations exist among molecular clouds of the Large Magellanic Cloud.  According to their fig.~6, $\eff$ ranges from $\simeq 1$ down to less than $10^{-3}$, the most massive clouds being the least efficient ones \citep[see also][]{mur11}.

This debate regarding the variations of the \sfe per free-fall time or the absence thereof will not move forward unless it takes into account the impact that the structure of \sfing regions has on their \stf activity.  Not only does the \sfr depend on the \sfe per free-fall time, on the mass and \fft of the \sfing gas, it also depends on how centrally-concentrated the gas is, an aspect which remains unaccounted for by Eq.~\ref{eq:intro}.  

In this contribution, we expand the cluster-formation model of \citet{par13} to estimate the ratio between the \sfr of centrally-concentrated clumps and the \sfr that they would have if they were of uniform density.  We refer to this ratio as the {\it magnification factor}, this one encapsulating the impact of the structure of a clump on its \stf rate.  
We also obtain the \sfe per \fft of clumps that one would measure based on their global properties (\stf rate, gas mass and free-fall time), that is, using Eq.~\ref{eq:intro} and ignoring their degree of concentration.  We are then able to contrast this {\it measured} \sfe per \fft with its counterpart for a clump region small enough to be considered of uniform density, which we refer to as the {\it intrinsic} \sfe per free-fall time (that is, by comparing the intrinsic and measured \stf efficiencies per free-fall time, we compare a local quantity with a global one).  This will yield a map of how distinct {\it intrinsic} and {\it measured} \stf efficiencies per \fft are, as a function of other clump properties. 

In the simulations presented here, the intrinsic \sfe per \fft is assumed to be constant, both in space and in time. 
Our model \sfing regions are spherically-symmetric with radii ranging from a fraction of a parsec up to $\simeq 10$\,pc.  We coin them "clumps", while stressing that other authors may use the term "cloud" in similar contexts.   

The outline of the paper is as follows.  Section \ref{sec:mod} introduces the concepts of intrinsic and measured \stf efficiencies per free-fall time, and the concept of magnification factor.  Section \ref{sec:obs} presents observational evidence from nearby molecular clouds and Sec.~\ref{sec:prelim} presents the model.  Section \ref{sec:dg} maps the magnification factor as a function of clump mass, radius, density profile and time.  In Sec.~6, we briefly discuss two model consequences.  
Conclusions are presented in Sec.~\ref{sec:conc}. 

\section{Measured vs. Intrinsic Star Formation Efficiency per Free-Fall Time}
\label{sec:mod}
Applying Eq.~\ref{eq:intro} to a molecular clump with known estimates of its gas mass, $m_{gas}$, free-fall time, $\langle\tff\rangle$, and \stf rate, $SFR_{clump}$, one can derive a \sfe per free-fall time: 
\begin{equation}
\effmeas = \frac{SFR_{clump}\langle\tff\rangle}{m_{gas}}\,,
\label{eq:effmeas}
\end{equation} 
which we refer to as the {\it measured} \sfe per free-fall time. 
In Eq.~\ref{eq:effmeas}, the free-fall time $\langle\tff\rangle$ is calculated at the mean volume density $\langle \rho_{gas} \rangle$ of the clump gas:
\begin{equation}
\langle \tff \rangle = \sqrt{\frac{3\pi}{32G\langle \rho_{gas} \rangle}}\,,
\label{eq:tffglob}
\end{equation}
and $\effmeas$ is therefore a global value.

Although Eq.~\ref{eq:effmeas} does not explicitly account for the structure of the clump, it is nevertheless affected by it.  
To reveal this effect, we compute the total \sfr of the clump by integrating the \sfr of spherical shells of gas, from clump center to clump edge: 
\begin{equation}
\begin{aligned}
SFR_{clump} & = \int_0^{r_{clump}} \effint \frac{dm_{gas}(r)}{\tff(r)}  \\
                      & =  \int_0^{r_{clump}} \effint \frac{4 \pi r^2 \rho_{gas}(r)}{\tff(r)}dr \;.
\end{aligned}
\label{eq:sfrint}
\end{equation}
Here, $r_{clump}$ is the clump radius, $dm_{gas}(r) = 4 \pi r^2 \rho_{gas}(r) dr$ is the gas mass of a spherical shell of radius $r$,  thickness $dr$, gas volume density $\rho_{gas}(r)$ and \fft $\tff(r)$, with $0\leq r \leq r_{clump}$.  In contrast to Eq,~\ref{eq:tffglob}, $\tff(r)$ is a locally-defined free-fall time, i.e.:
\begin{equation}
\tff(r) = \sqrt{\frac{3\pi}{32G \rho_{gas}(r)}}\;.
\label{eq:tffloc}
\end{equation}
We refer to $\effint$ as the {\it intrinsic} \sfe per free-fall time, namely, the efficiency characterizing the \stf activity of nested shells of gas, each shell being narrow enough to have its own uniform volume density $\rho(r)$ and \fft $\tff(r)$.  We assume that $\effint$ is constant, varying neither as a function of radial location inside the clump, nor as a function of time. 

For a pure power-law density profile of slope $-p$ $\rho_{gas}(r) \propto r^{-p}$, Eq.~\ref{eq:sfrint} yields (for $p<2$):
\begin{equation}
SFR_{clump} = \frac{(3-p)^{3/2}}{2.6(2-p)} SFR_{TH}\,,
\label{eq:sfrpl}
\end{equation}  
where
\begin{equation}
SFR_{TH} = \effint \frac{m_{gas}}{\tff}
\label{eq:sfrth}
\end{equation}   
is the \sfr that the clump would have were its density profile a top-hat one (i.e. $SFR_{clump}=SFR_{TH}$ when $p=0$).  In this case, the local \fft $\tff(r)$ is independent of $r$ and equates its global counterpart $\langle\tff\rangle$.    
Equation \ref{eq:sfrpl} is but for the numerical factor 2.6 in the denominator identical to equation 2 in \citet{tan06}, where the numerical factor is 2.3.  The difference between both stems from \citet{tan06} accounting for variations of $\effint$ with the gas Mach number, while $\effint$ is considered constant in Eq.~\ref{eq:sfrint}.  
  
Comparing Eq.~\ref{eq:sfrth} to Eq.~\ref{eq:intro}, one can see that if $\eff$ in Eq.~\ref{eq:intro} is the intrinsic \sfe per free-fall time, Eq.~\ref{eq:intro} can only be applied to gas reservoirs of uniform density.  Conversely, if Eq.~\ref{eq:intro} is applied to any \sfing region, regardless of its structure, $\eff$ is necessarily the globally-measured \sfe per free-fall time (Eq.~\ref{eq:effmeas}).  Equation~\ref{eq:sfrpl} therefore provides a more reliable parameterization of the clump \sfr as it disentangles the respective contributions of the intrinsic \sfe per \fft and of the clump density profile.  That is:  
\begin{equation}
SFR_{clump} =  \zeta \cdot SFR_{TH} = \zeta \cdot \effint \frac{m_{gas}}{\tff}\,,
\label{eq:sfrzeta1}
\end{equation}  
where $\zeta$ is the magnification factor, namely, the factor by which the slope of the density profile of the clump enhances its \stf rate compared to the case of a top-hat profile.  
Comparing Eq.~\ref{eq:sfrzeta1} with Eq.~\ref{eq:effmeas} also shows that the measured \sfe per \fft is the product of its intrinsic counterpart and of the magnification factor $\zeta$, that is:
\begin{equation}
\effmeas = \zeta \effint \;.
\label{eq:effdeg}
\end{equation} 
Equation \ref{eq:effdeg} thus encapsulates  the degeneracy existing between the intrinsic \sfe per \fft $\effint$ on the one hand, and the impact $\zeta$ of the clump density profile on the other hand.

\section{Hints from nearby clouds}
\label{sec:obs}

\begin{figure}
\begin{center}
\epsscale{1.0}  \plotone{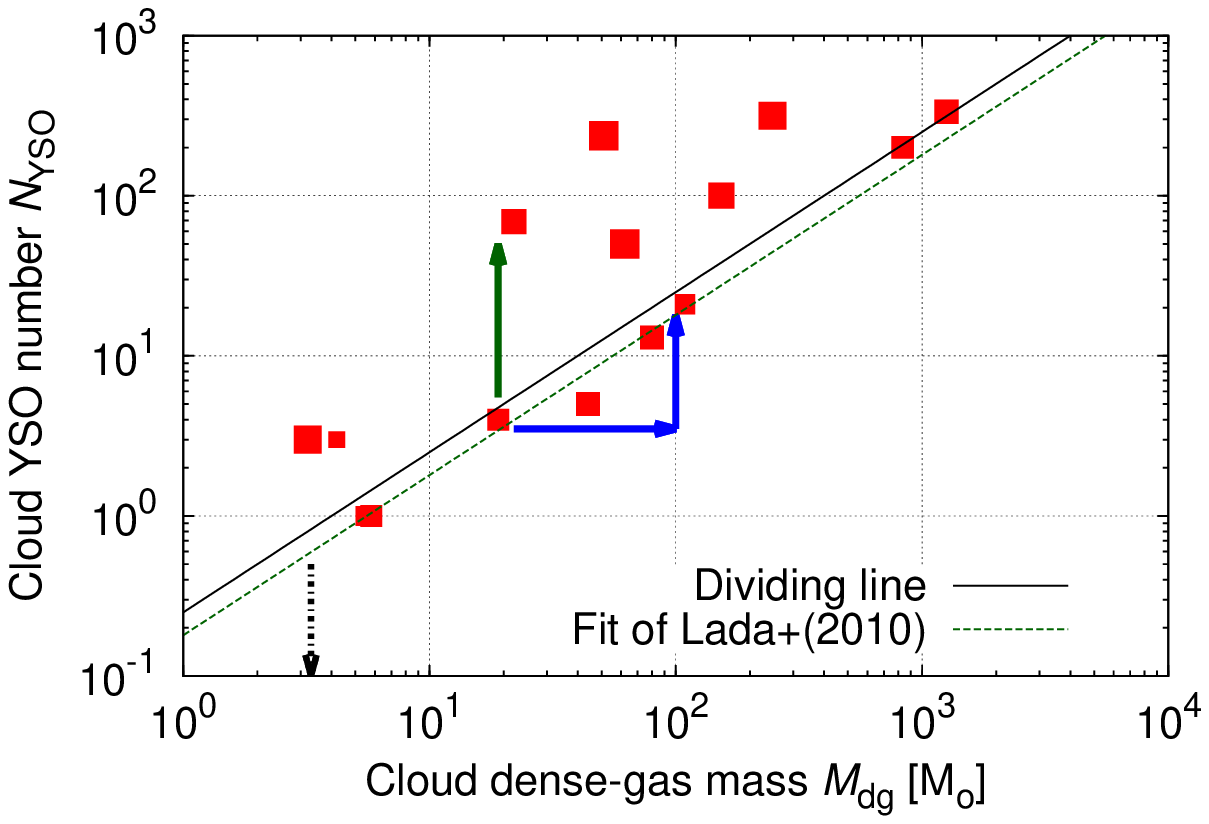}
\caption{Relation between the dense gas mass and the number of YSOs of nearby molecular clouds.  The downward dashed arrow indicates a starless cloud.  Larger symbols code steeper gas density profiles ($p$ is observed to range from $1.17$ to $2.05$; data taken from table~S1 in \citet{kai14}).  The solid line is our dividing line, defined with a slope of $1$ as for the fit of \citet{lad10}.  It divides the cloud sample in two equal-size subsamples, with respective average values for $p$ of $\langle p_{above} \rangle = 1.81\pm0.27$ and $\langle p_{below} \rangle = 1.51\pm0.14$.  While it is commonly accepted that a cloud increases its YSO census by increasing its dense-gas mass (blue arrows), the data above suggest that it may also do so by steepening its density profile ($p$-increase: upward green arrow).      }
\label{fig:tst2}
\end{center} 
\end{figure}

Figure~\ref{fig:tst2} shows the correlation between the number of Young Stellar Objects (YSOs) and the mass of dense gas of a sample of nearby molecular clouds studied by \citet{kai14} (data from their table~S1).  It is thus akin to fig.~4 of \citet{lad10}, whose linear fit is shown by the green dashed line.  We stress, however, that \citet{lad10} and \citet{kai14} do not define the dense gas in the same way.  While \cite{lad10} define the dense gas as gas denser than a surface density $\Sigma_{gas}=160\,\Mspp$, \citet{kai14} define it based on a volume/number density threshold of $\simeq 5 \cdot 10^3\,\cc$.  This explains why the fit from \citet{lad10} overestimates the dense gas mass of most clouds: their (projected) dense gas mass includes foreground and background contributions with respect to what is measured by \citet{kai14} \citep[see also Fig.~11 in ][]{par13}.  Figure~\ref{fig:tst2} also depicts the steepness $p$ of the density profile of each cloud as the symbol size, larger symbols corresponding to steeper profiles.  $p$ has been inferred from the slope of the gas density-PDF and is given as the $\kappa$ parameter in table~S1 of \citet{kai14}.  The density profile steepness being measured over the gas density range for which the cloud is structured \citep[i.e. the gas density-PDF is a power law; see e.g.][]{kri11}, the steepness $p$ "covers" the dense gas (up to the spatial resolution limit; see below) but also some gas at lower volume densities.  

Figure \ref{fig:tst2} shows that steeper profiles tend to top the data compared to shallower profiles.  To ascertain this, we have divided the data points in two groups, located above, respectively, under, a dividing line of slope 1 whose intercept is such that there are 8 clouds in each group.  Our dividing line is very similar to the linear fit of \citet{lad10}.  Clouds above and below the dividing line have a mean steepness of $\langle p_{above} \rangle = 1.81\pm0.27$ and $\langle p_{below} \rangle = 1.51\pm0.14$, respectively.  This suggests that the scatter of the $N_{YSO}$-$M_{dg}$ relation is not entirely random, and is partly driven by the gas density gradient.  That is, to increase its YSO production, a cloud can either increase its dense-gas mass, or steepen the gas density profile.  This is as anticipated by the model presented in Sec.~\ref{sec:mod}.  

When considering Fig.~\ref{fig:tst2}, however, one should keep in mind the following four points: 
(i) not all clouds are at the same evolutionary stage, and a higher YSO census may also stem from a longer \stf episode; 
(ii) it is unclear which fraction of their YSOs clouds actually form in their dense gas as the existence of a gas density threshold for \stf remains debated \citep{par16,par17,elm18}; 
(iii) $p$ is unknown in the number density regime beyond the resolution limit of the observations, i.e. $>5 \cdot 10^4\,\cc$ \citep{kai14}; 
(iv) the observed YSO census measures the {\it past} \stf history and, therefore, results from the cloud and clump structures prevailing {\it before} the time $t$ of the observations; presently-observed structures, as quantified by the observed steepness $p$, drive the {\it present and forthcoming} \stf rate.  The coupling between the present-day density gradient and the past \stf rate can thus never be complete since both quantities pertain to different time-spans.

Figure~\ref{fig:tst2} nevertheless constitutes a strong incentive to study the interplay between clump structure and \stf rate, especially for the uncharted regime $p>2$, where an impact significantly stronger than for $p<2$ is expected.   
   
\section{Preliminary Insights into the Model}
\label{sec:prelim}
\begin{figure*}
\begin{center}
\epsscale{1.0}  \plotone{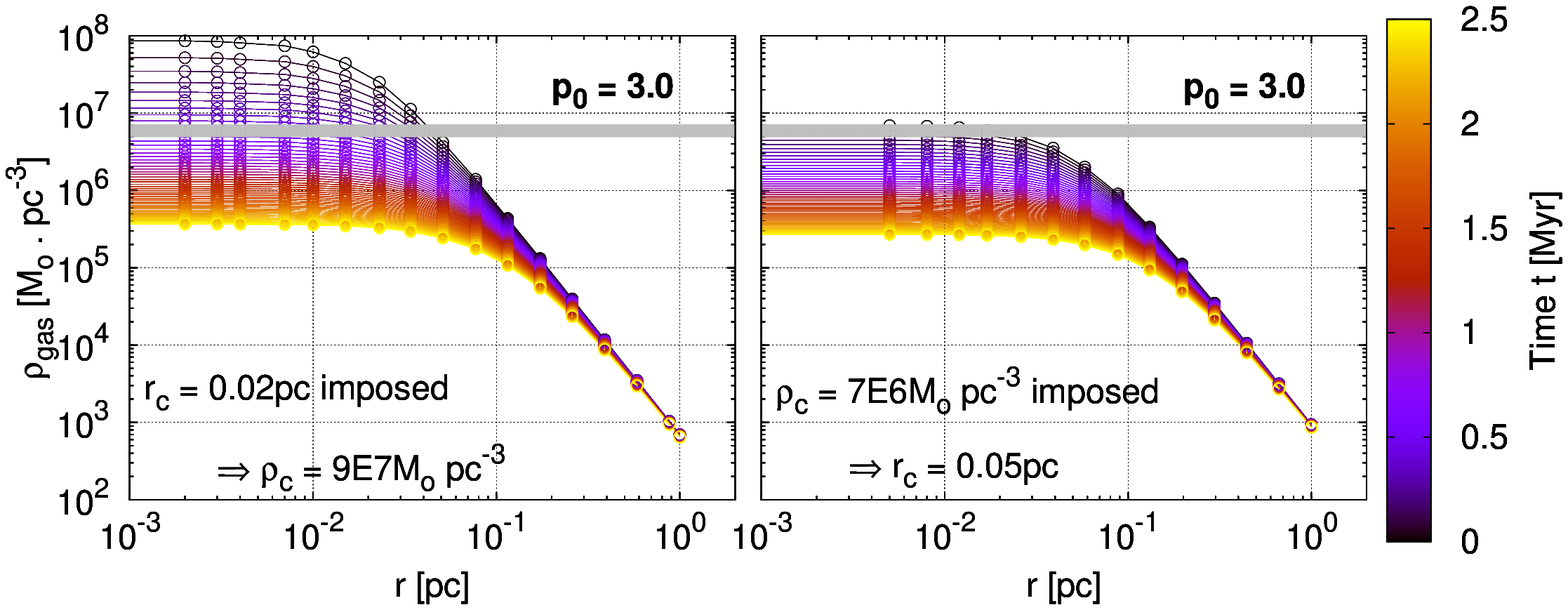}
\caption{Evolution with time of the gas density profile of two model clumps.  Common model parameters are $m_{clump} = 3.2\cdot10^4\,\Ms$, $r_{clump} = 1$\,pc, $p_0=3$ and $\effint=0.01$.  The horizontal grey stripe depicts the density of the densest known protostars (i.e. $ \rho \simeq 7\cdot10^6\,\Msppp$).  {\it Left panel:}  Initial core radius imposed, $r_c=0.02$\,pc, and initial central density $\rho_c$ inferred.  {\it Right panel:} Initial gas central density imposed, $\rho_c = 7\cdot10^6\,\Msppp$, and core radius $r_c$ inferred.  The color-coding corresponds to time $t$ after star formation onset, from 0\,Myr to 2.5\,Myr.  }
\label{fig:2prof}
\end{center} 
\end{figure*}

We need to numerically integrate Eq.~\ref{eq:sfrint} to compute $SFR_{clump}$ and to map the magnification factor
\begin{equation}
\zeta = \frac{SFR_{clump}}{SFR_{TH}} = \frac{\effmeas}{\effint}
\label{eq:zeta1}
\end{equation}
as a function of clump density profile, mass, radius and time.  We assume that, at the onset of \stf (i.e. $t=0$), the gas  density profile of clumps obeys a power-law with a central core of size $r_c$ so as to avoid a density singularity at the clump center:
\begin{equation}
\rho_{gas} (t=0, r) = \frac{\rho _c}{\left[1+(\frac{r}{r_c})^2\right]^{p_0/2}}\;.
\label{eq:denpro}
\end{equation}
In this equation, $\rho_c$ is the initial clump central density and $-p_0$ is the slope of the initial clump density profile when $r>>r_c$.  Both $p_0$ and $r_c$ control how centrally-condensed a clump  initially is and, therefore, also control its \sfr $SFR_{clump}$ and magnification factor $\zeta$.  All models are given an intrinsic \sfe per \fft  $\effint=0.01$. 
A model clump is thus defined by its intrinsic \sfe per free-fall time, its mass $m_{clump}$ enclosed within its radius $r_{clump}$ ($m_{clump}$ is equivalent to the initial gas mass), the initial steepness $p_0$ of its density profile and, either its initial core radius $r_c$, or its initial central density $\rho_c$, one imposing the other such that the mass enclosed within the radius $r_{clump}$ is the clump mass $m_{clump}$.  We run two categories of models.  Firstly, we impose an initial core radius and calculate the initial gas central density $\rho_c$ such that a mass $m_{clump}$ is enclosed within the clump radius $r_{clump}$ given the adopted density profile (Eq.~\ref{eq:denpro}).  We adopt  $r_c=0.02$\,pc, which is of order the size of the circumstellar envelope of a protostar \citep{mot98}.  The initial central density $\rho_c$ of such models can be extremely high when they are both dense and very centrally concentrated.  An example is shown in the left panel of Fig.~\ref{fig:2prof}.  Yet, the densest known protostars have a number density $n_{H2} \simeq 10^8\,\cc$ \citep{mot18}, equivalent to a mass density $\rho \simeq 7\cdot10^6\,\Msppp$.  
Therefore, we shall also discuss how the results for $p_0=3$ and $p_0=4$ are affected when imposing an initial central density $\rho_c = 7\cdot10^6\,\Msppp$ instead of imposing the initial core radius $r_c$.  The panels of Fig.~\ref{fig:2prof} compare the evolution with time of both types of density profiles for otherwise identical model parameters.  The gas density profile at time $t>0$ is given by eqs~18 and 19 of \citet{par13} (equations in which $\rho_0(r)$ is the initial gas profile).    
We note that densities higher than $\rho_c = 7\cdot10^6\,\Msppp$ may nevertheless be relevant to specific topics, e.g. for the progenitors of the supermassive stars invoked by \citet{den14} to explain the light-element abundance anomalies of old globular clusters.
 
\begin{figure}
\begin{center}
\epsscale{1.0}  \plotone{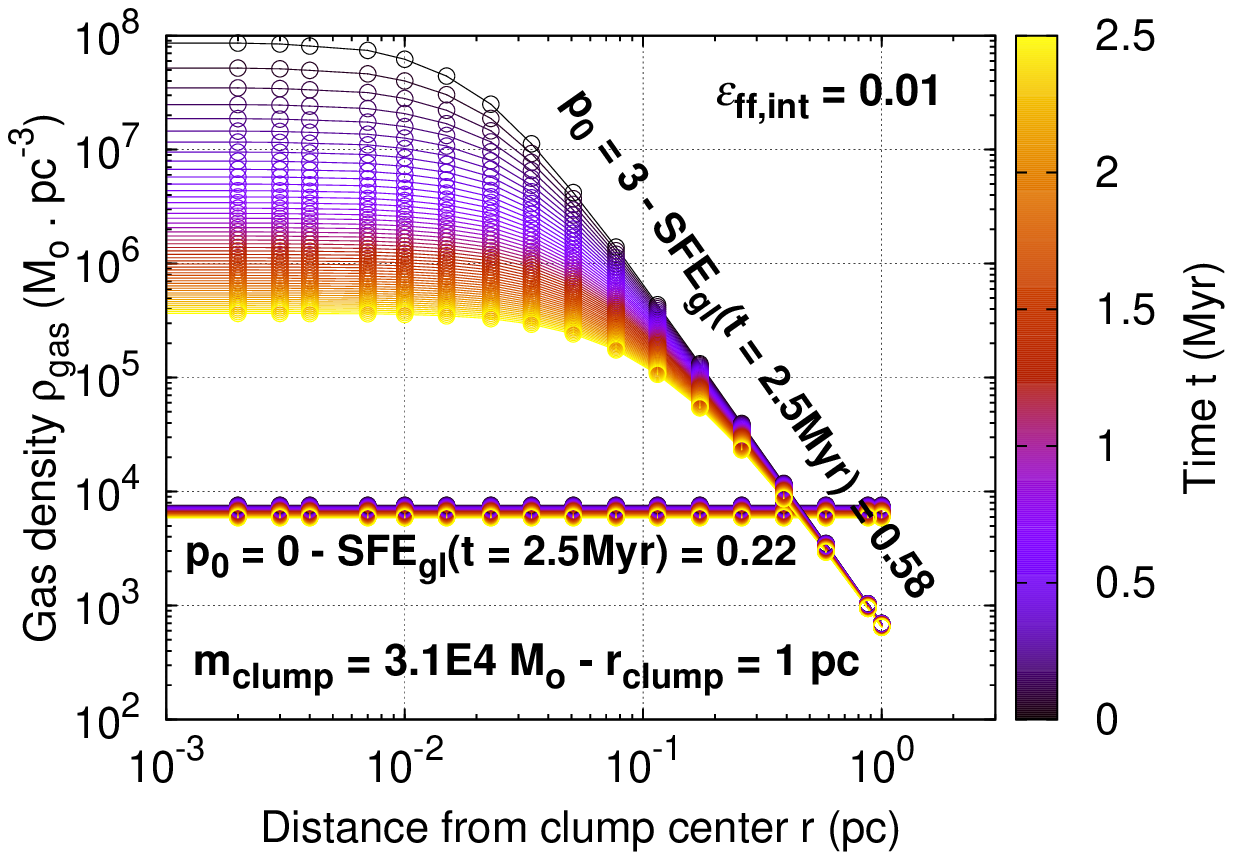}
\caption{Evolution with time of the gas volume density profile of two model clumps with an intrinsic \sfe per \fft $\effint = 0.01$, an initial gas mass $m_{clump} \simeq 3.2\cdot10^4\,\Ms$ and a radius $r_{clump}=1$\,pc.  In one case, the gas density profile is initially a top-hat profile, and stays so through the full course of the simulations.  In the other case, the parameters of the initial gas density profile are $p_0=3$ and $r_c=0.02$\,pc (see Eq.~\ref{eq:denpro}).  The color of each profile gives the time elapsed since star formation onset (see the palette at the right-hand side).}
\label{fig:denprotim}
\end{center} 
\end{figure}    

As we shall map how the magnification factor evolves as a function of time, let us consider how the \sfr evolves with time in this class of models.
As time goes by, the mass of the gas reservoir dwindles, its volume density decreases and its \fft increases, all factors yielding a decreasing \stf rate \citep{par14}.  The inner regions of a centrally-concentrated clump, due to their high density/short free-fall time, are those experiencing the fastest decrease of their gas volume density as their gas is being fed to forming stars (see Fig.~\ref{fig:2prof}).  This is illustrated further in Fig.~\ref{fig:denprotim}, which compares the evolution with time of a gas density profile with $p_0=3$ with that of a top-hat model ($p_0=0$) with identical mass and radius ($m_{clump} \simeq 3.2\cdot10^4\,\Ms$, $r_{clump}=1$\,pc).  Both models thus differ only in their initial density profile.  
The \stf time-span is color-coded by the right-hand-side palette.  The global \stf efficiency, $SFE_{gl}$ (i.e. the clump gas mass fraction turned into stars), achieved at any given time is significantly higher for $p_0=3$ than for $p_0=0$.  As an example, Fig.~\ref{fig:denprotim} gives $SFE_{gl}$ at $t=2.5$\,Myr for both cases.  This illustrates once again that the activity of a \sfing region gets boosted simply by steepening its density profile.  Since, in a centrally-concentrated clump, the inner regions are those boosting the \stf rate, once their density  dwindles, the clump overall \sfr decreases as well, with initially steeper  density profiles being conducive to sharper decreases with time of the \stf rate.

\begin{figure}
\begin{center}
\epsscale{1.0}  \plotone{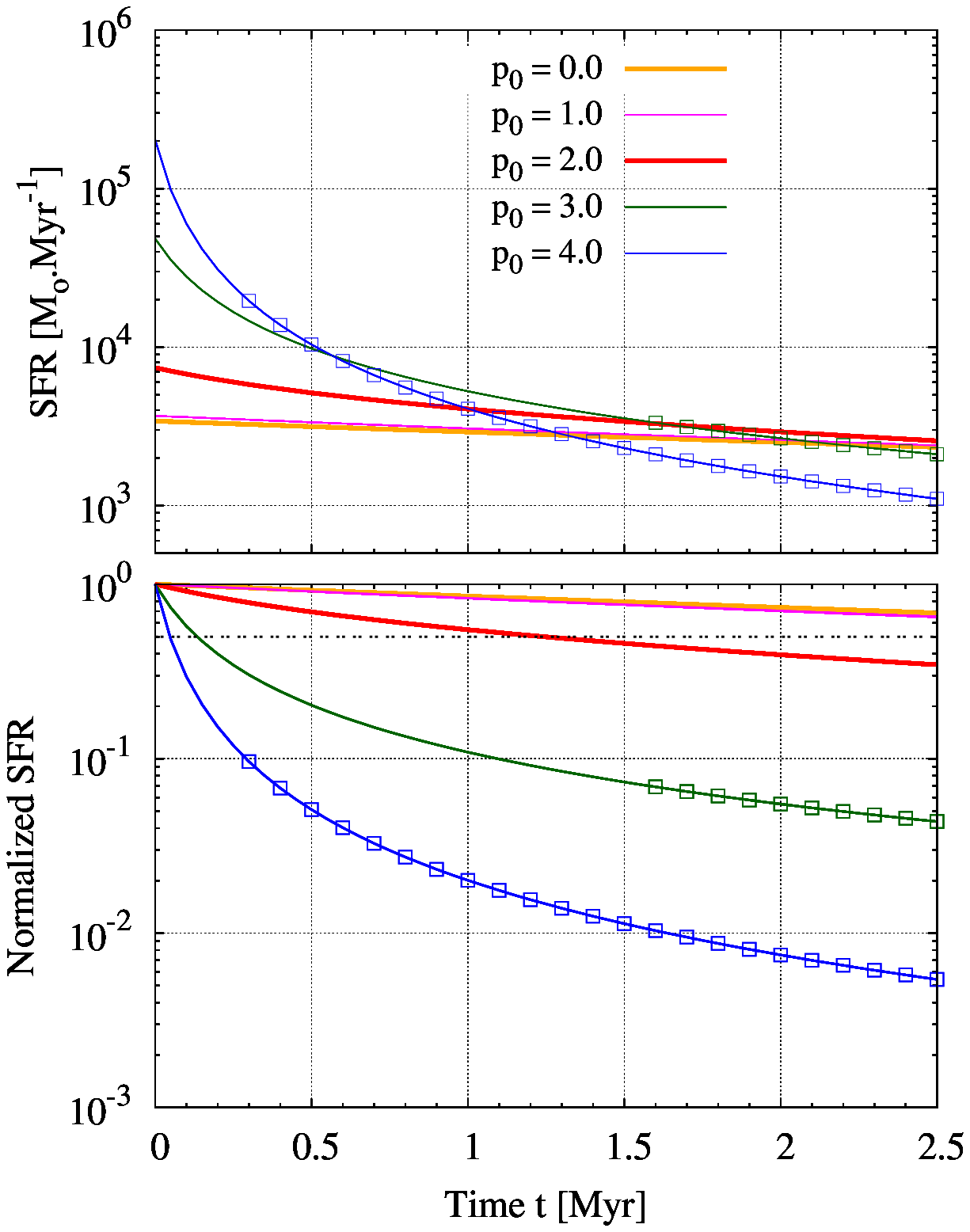}
\caption{Top panel: Evolution with time of the \sfr SFR of five model clumps, each with a different steepness $p_0$ of the initial gas density profile (see the key).  Other parameters ($\effint$, $m_{clump}$, $r_{clump}$ and $r_c$) are identical to those used in Fig.~\ref{fig:denprotim}.  Bottom panel: Same as in top panel, but with the \sfr normalized to its initial value.  The black horizontal dotted line indicates when the \sfr has dropped to half its initial value.  In both panels, line segments with open squares indicate the time-spans over which the global \sfe $SFE_{gl}$ has become higher than 0.5 (which happens for $p_0=3$ and $p_0=4$ only).   While the top panel highlights that a steeper gas density profile promotes a higher \sfr initially, the bottom panel shows that it also yields a faster decline of the \stf rate with time. }
\label{fig:sfh_2p}
\end{center} 
\end{figure}

\begin{figure*}[t]
\begin{center}
\epsscale{1.0}  \plotone{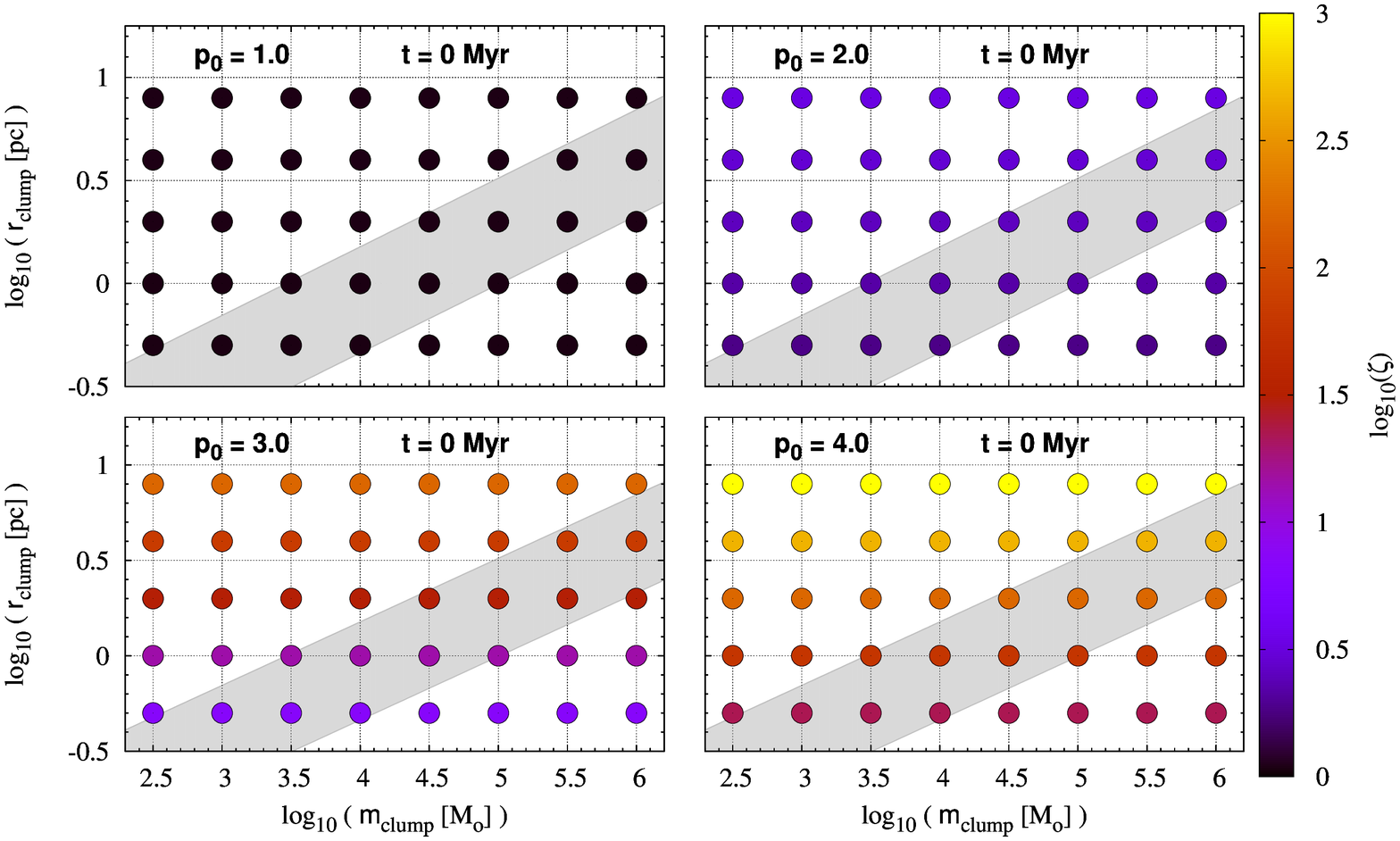}
\caption{Magnification factor $\zeta$ of the \sfr of model clumps at \stf onset (i.e. $t=0$\,Myr), with the initial core radius imposed.   Parameters are the initial gas mass and radius of clumps, $m_{clump}$ and $r_{clump}$, and the initial steepness $p_0$ of their density profile.   {\it Top-right panel:} $p_0=1$.  {\it Top-left panel: } $p_0=2$.  {\it Bottom left-panel:}  $p_0=3$.  {\it Bottom-right: } $p_0=4$.  The intrinsic \sfe per \fft and the core radius of the initial gas density profile are set to $\effint = 0.01$ and $r_c = 0.02$ pc for all simulations.  Each model is represented by a plain symbol, the color of which depicts the value of $\zeta$ (see palette for color-coding, with a logarithmic scale ranging from $\zeta = 1$ to $\zeta = 10^3$).  The grey stripe highlights the density regime for which steep density profiles have been detected in Galactic clouds \citep[i.e. $10^4\,\cc < n_{H2} < 3\cdot10^5\,\cc \equiv 700\,\Msppp < \langle \rho_{clump} \rangle < 2.1\cdot10^4\,\Msppp$;][]{schn15}.   }
\label{fig:map1}
\end{center} 
\end{figure*}

This is illustrated in the top panel of Fig.~\ref{fig:sfh_2p}, which presents the evolution with time of the \sfr of 5 model clumps with the same intrinsic \sfe per free-fall time, mass, radius and central core radius as in Fig.~\ref{fig:denprotim}, but 5 different steepnesses of the density profile, from $p_0=0$ to $p_0=5$ in steps of 1.  The line segments with open squares indicate the time-spans over which the global \sfe $SFE_{gl}$ has become higher than 0.5 (for $p_0=3$ and $p_0=4$).  

The evolution with time of the $p_0=0$ and $p_0=1$ models are very similar (compare the orange and magenta lines).  Once the density profile gets steeper, however, not only does the \sfr at \stf onset get higher, the time-scale on which the \sfr decreases gets shorter too.  This is clearly illustrated by the bottom panel of Fig.~\ref{fig:sfh_2p}, which shows the \sfr normalized to its initial value.  While the model with $p_0=4$ loses half of its initial \sfr in less than 0.1\,Myr, the models with $p_0=0$ and $p_0=1$ need more than 2.5\,Myr to do so.  The \sfr for $p_0=4$ decreases so fast that, in theory, it could become smaller than the \sfr for $p_0=0$ (see top panel, where the blue line eventually finds itself below the orange one).  For the model parameters used here and $p_0=4$, this is predicted to happen at $t \simeq 1.2$\,Myr, when the \sfe is already $SFE_{gl}=0.77$, a case calling for gas exhaustion -- rather than gas expulsion -- in \sfing regions \citep{wat19}.

\section{Mapping the Magnification Factor $\zeta$}
\label{sec:dg}

In this section, we map, and we understand, the variations of the magnification factor $\zeta$ as a function of clump parameters and time.  We impose firstly an initial core radius $r_c$, secondly an initial central density $\rho_c$.  The radius and mass of our model clumps extend from $r_{clump} = 0.5$\,pc to 8\,pc in logarithmic steps of 0.30, and from $m_{clump} = 300\Ms$ to $10^6\,\Ms$ in logarithmic steps of 0.50.  We remind that $m_{clump}$ is the clump initial gas mass enclosed within $r_{clump}$.

\subsection{$r_c$ imposed ($r_c=0.02$\,pc) and $\rho_c$ inferred}

\begin{figure*}
\begin{center}
\epsscale{1.0}  \plotone{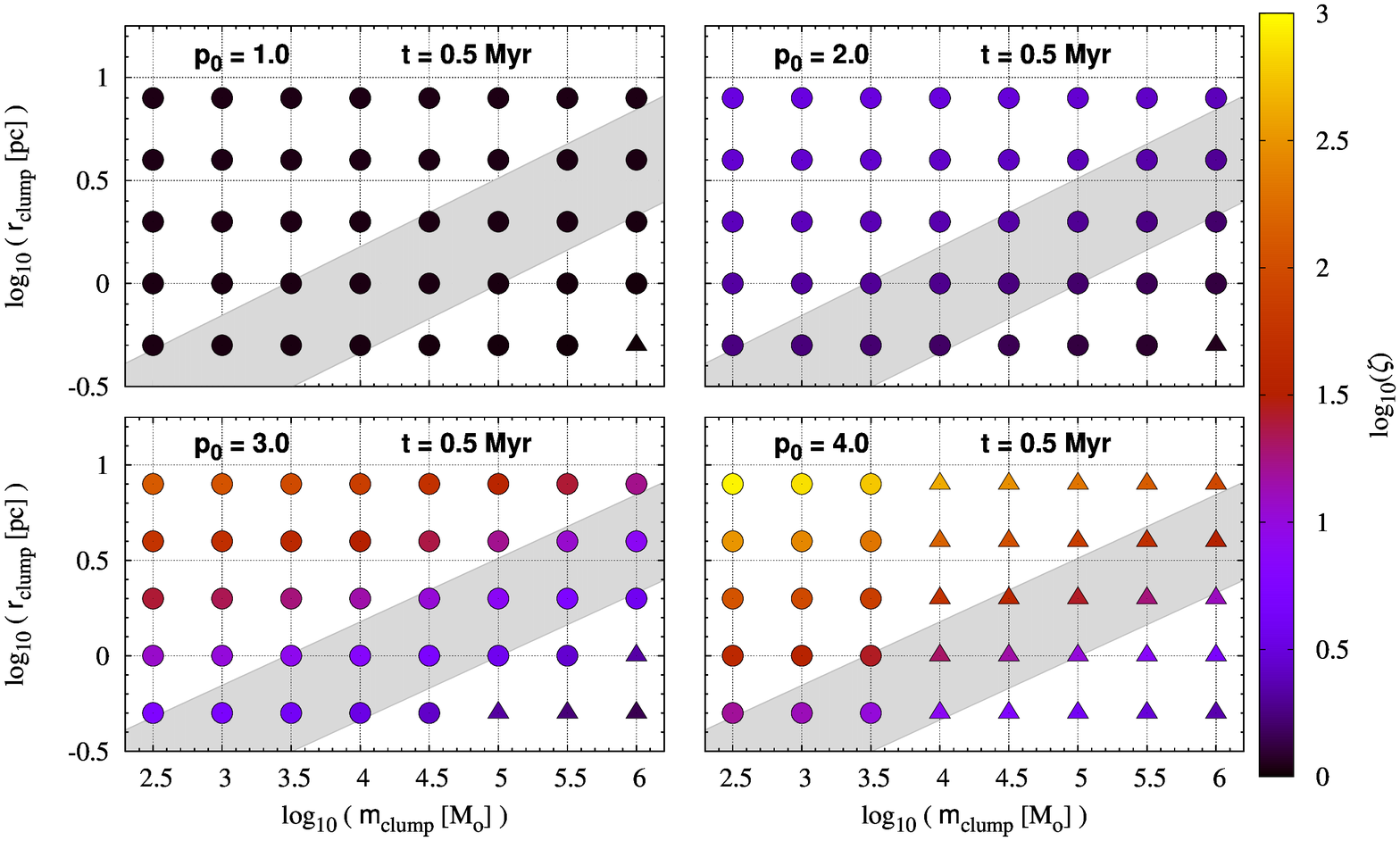}
\caption{Same as in Fig.~\ref{fig:map1}, but $t=0.5$\,Myr after the onset of star formation.  Model clumps which by this \stf time-span have achieved $SFE_{gl}>0.5$ are depicted by triangles}
\label{fig:map2}
\end{center} 
\end{figure*}

Results are presented in Fig.~\ref{fig:map1} for \stf onset ($t=0$\,Myr) and $r_c=0.02$\,pc.  Each panel corresponds to one value of the clump profile steepness, from $p_0=1$ (top-left panel) to $p_0=2$ (top-right panel), $p_0=3$ (bottom-left panel) and $p_0=4$ (bottom-right panel).    Each model clump is depicted by a plain circle whose color codes its magnification factor $\zeta$ according to the right-hand-side color palette.  It should be noted that the palette gives $log_{10}\zeta$.  The maximum $\zeta$ value obtained here is $\simeq 1,300$ (for $p_0=4$ and $r_{clump}=8$\,pc).  We caution that not the entire parameter space may be physically relevant.  For instance, the top-left corner of each panel consists of clumps with a mean volume density lower than that of Galactic molecular clouds.  We nevertheless include such models in the discussion so as to  understand the variations of $\zeta$ through the full extent of the parameter space.    
    
When $p_0=1$, clumps are barely centrally-concentrated, and their \sfr increases by less than $\simeq 10$\,\% with respect to their counterpart with $p_0=0$.  For $p_0=2$, the increase becomes more noticeable, with $\zeta \simeq 2{\rm -}3$.

As the profile steepness reaches $p_0=3$ and $p_0=4$, $\zeta$ truly deserves to be coined the "{\it magnification factor}".  The \sfr is indeed boosted by about two ($p_0=3$) and three ($p_0=4$) orders of magnitude when the ratio between the core radius and the clump radius is at its smallest value (i.e. $r_c/r_{clump}=0.0025$ when $r_{clump}=8$\,pc; top row of symbols in each panel).  As for the clumps with the largest $r_c/r_{clump}$ ratio (i.e. $r_c/r_{clump}=0.04$ when $r_{clump}=0.5$\,pc; bottom row of symbols), the magnification factor reaches smaller, albeit still significant, values: $\zeta \simeq 7$ ($p_0=3$) and $\zeta \simeq 20$ ($p_0=4$).  

For a given steepness $p_0$ of the initial density profile, clumps with the largest radius experience therefore the greatest \sfr increase with respect to a homogeneous model.  This is because their density profile is the most centrally-peaked, i.e. their $r_c/r_{clump}$ ratio is the smallest (recall that $r_c$ is kept constant).  The parameterization of the clump density profile should thus not be reduced to its initial slope $-p_0$. The relative extent $r_c/r_{clump}$ of the central core matters as well since it contributes to the central-peakedness of the clump density profile.  If all models had the same $r_c/r_{clump}$ ratio -- rather than a given core radius $r_c$ -- all symbols of any given panel would have the same color.  That is, the value of $\zeta$ would be independent of both $m_{clump}$ and $r_{clump}$.  To summarize, at $t=0$, the magnification factor $\zeta$ depends on both $p_0$ and $r_c/r_{clump}$, i.e. $\zeta(t=0) = \zeta(p_0,r_c/r_{clump})$.    
  
That the measured \sfe per \fft gets higher for smaller $r_c/r_{clump}$ ratios is reminiscent of the results obtained by \citet{cho11}.  For a given gas mass (i.e. the mass of their computational box), the factor by which their core formation efficiency per \fft is enhanced when turning-on self-gravity is higher for greater density ratios $\rho_{crit}/\langle \rho_0 \rangle$, with $\rho_{crit}$ the adopted critical density for star formation and $\langle \rho_0 \rangle$ the mean volume density of the simulated gas.  The increasing factors are 2.2 and 2,400 for $\rho_{crit}/\langle \rho_0 \rangle = 30$ and $\rho_{crit}/ \langle \rho_0 \rangle = 500$, respectively.  Under the assumption of a constant critical density for star formation, 
this yields a higher magnification factor when the volume density of the simulated gas is lower.  This is exactly as found in the semi-analytical calculations presented here since clumps of a given mass achieve greater magnification factors for lower mean volume densities (provided that $r_c$ remains constant or varies more slowly than $r_{clump}$).
    
\begin{figure*}[t]
\begin{center}
\epsscale{1.0}  \plotone{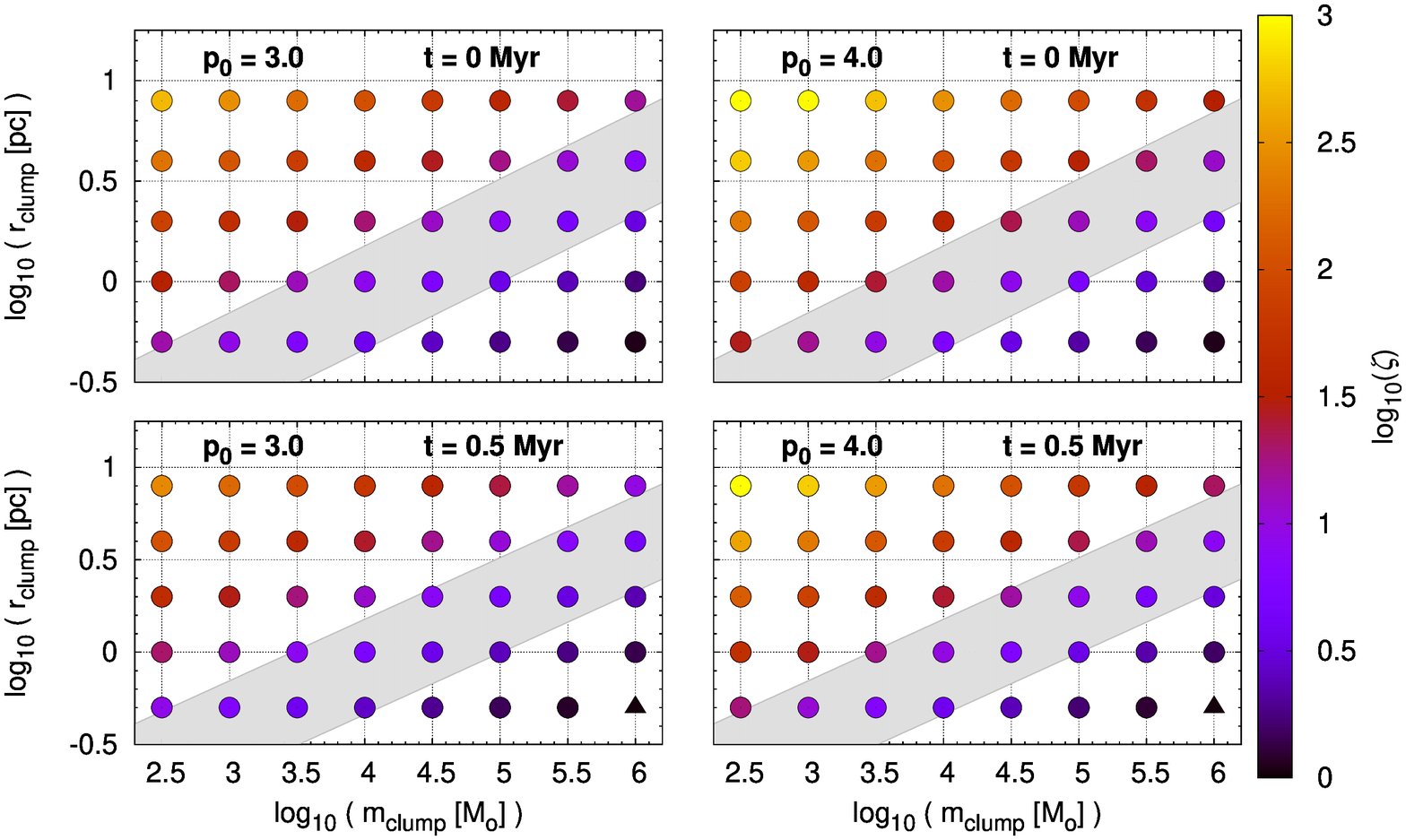}
\caption{
Magnification factor $\zeta$ of the \sfr of model clumps, with the initial central density imposed.  Parameters are the initial gas mass and radius of clumps, $m_{clump}$ and $r_{clump}$, and the initial steepness $p_0$ of their density profile.  {\it Top-left:} $p_0=3$ and $t=0$\,Myr.  {\it Top-right:} $p_0=4$ and $t=0$\,Myr.  {\it Bottom-left:} $p_0=3$ and $t=0.5$\,Myr.     {\it Bottom-right:} $p_0=4$ and $t=0.5$\,Myr.  The intrinsic \sfe per \fft and the initial gas central density are set to $\effint = 0.01$ and $\rho_c = 7\cdot10^6\,\Msppp$  for all simulations.  Each model is represented by a plain symbol, the color of which depicts the value of $\zeta$ (see palette for color-coding, with a logarithmic scale ranging from $\zeta = 1$ to $\zeta = 10^3$).  The grey stripe highlights the density regime for which steep density profiles have been detected in Galactic clouds (i.e. $10^4\,\cc < n_{H2} < 3\cdot10^5\,\cc \equiv 700\,\Msppp < \langle \rho_{clump} \rangle < 2.1\cdot10^4\,\Msppp$).   Model clumps which have achieved $SFE_{gl}>0.50$ by $t=0.5$\,Myr are depicted by triangles.}
\label{fig:rhoclim}
\end{center} 
\end{figure*}

As explained in Sec.~\ref{sec:prelim}, the frantic \stf activity shown by the $p_0=3$ and $p_0=4$ models at $t\gtrsim0$\,Myr cannot be sustained for long.  As the clump gas density profile loses its initial central peakedness, its \sfr and magnification factor start to wane. 
Figure \ref{fig:map2} maps the magnification factor at $t=0.5$\,Myr, showing that this one has decreased with time.  Clumps having achieved a global \sfe higher than $0.50$ are depicted by triangles.  Their ability to form stars at a fast pace results either from a high mean volume density (hence short free-fall time), or from a strongly centrally-peaked density profile (achieved either through a steep density profile, or through a smaller $r_c/r_{clump}$ ratio), or from a combination of these factors.  Note in that respect that the region of the diagram occupied by triangles is larger when $p_0=3$ or $p_0=4$ than when $p_0=1$ or $p_0=2$.  
Compared to Fig.~\ref{fig:map1}, $\zeta$ has decreased, as is especially noticeable when $p_0=4$.  This is so because, as we saw in the bottom panel of Fig.~\ref{fig:sfh_2p}, the \sfr decreases faster for steeper initial gas density profiles. 

For each combination of $p_0$ and $r_c/r_{clump}$, that is, for each row of symbols of each panel, the decrease with time of the magnification factor is stronger for more massive clumps since those have reached a more advanced stage of their evolution by virtue of their shorter mean free-fall time \citep[see also figs 3 and 4 in][]{par14}.  As in Fig.~\ref{fig:map1}, the highest value of $\zeta$ is achieved by the largest and least massive clump (top-left corner of each panel, corresponding to $m_{clump}\simeq300\,\Ms$ and $r_{clump}\simeq8$\,pc).  For $p_0=3$ ($p_0=4$), we still find for such clumps $\zeta \simeq 130$ ($\zeta \simeq 900$) when $t=0.5$\,Myr.  This stems from a high magnification factor initially, combined with a long mean \fft (e.g. $\langle \tff \rangle \simeq 20$\,Myr when $m_{clump}\simeq300\,\Ms$ and $r_{clump}\simeq8$\,pc), implying that the initially high magnification factor is sustained for a longer time-span than for denser clumps.  

In the density regime where steep density profiles have been detected (grey stripe), the most extreme variations of $\zeta$ are from $\simeq 20$ to $\simeq 700$ when $p_0=4$ and $t=0$.

\subsection{$\rho_c$ imposed ($\rho_c=7\cdot10^6\,\Msppp$) and $r_c$ inferred}

So far, we have imposed $r_c=0.02$\,pc and calculated the initial gas central density $\rho_c$ so that a mass $m_{clump}$ is enclosed within a radius $r_{clump}$ given the adopted density profile (Eq.~\ref{eq:denpro}).  $\rho_c$ can thus be extremely high for models which are both dense and very centrally concentrated, such as in Fig.~\ref{fig:denprotim} where $\rho_c \simeq 10^8\,\Ms \cdot pc^{-3}$.  Yet, the densest protostars have a number density $n_{H2} \simeq 10^8\,\cc$, equivalent to a volume $\rho \simeq 7\cdot10^6\,\Msppp$ \citep[table~2 in ][]{mot18}.  We now discuss how the results for $p_0=3$ and $p_0=4$ are affected when imposing a clump initial central density $\rho_c=7\cdot10^6\,\Msppp$ rather than a core radius $r_c=0.02$\,pc.  

We first stress that this is {\it not} the high central density {\it per se} which drives the initially high magnification factor when $p_0=3$ and $p_0=4$.  Should it be the case, $\zeta$ would vary with clump mass at a given clump radius in Fig.~\ref{fig:map1} since the initial central density $\rho_c$ increases with increasing clump mass when the shape of the profile ($r_c$, $p_0$) is fixed.  Rather this is the {\it density contrast} $\rho_c/\rho_{edge}$ between clump centre and clump edge which drives the value of $\zeta$ \citep[see also][]{elm11}.  This density contrast depends on the clump radius expressed in units of the core radius and on the steepness $p_0$ as $\rho_c/\rho_{edge}(t=0) \simeq (r_{clump}/r_c)^{p_0}$ (see Eq.~\ref{eq:denpro}).  These two parameters, $r_{clump}/r_c$ and $p_0$, were indeed identified in the previous section as those driving the initial magnification factor.  

Results are presented in Fig.~\ref{fig:rhoclim}: $p_0=3$: left panels; $p_0=4$: right panels; $t=0$\,Myr: top panels; $t=0.5$\,Myr: bottom panels.  Massive and compact clumps, to accommodate their large mass inside a given volume with a limited central density, adopt a fairly large core radius.  This strongly limits their initial magnification factor in spite of the steep density profile (e.g. $r_{clump}=1$\,pc, $m_{clump}=10^6\,\Ms$, $p_0=4$: $r_c=0.28$\,pc and $\zeta \simeq 2$).  This is also the reason why the initial magnification factor decreases with increasing clump mass for a given clump radius and a given density profile slope (i.e. given row in a given panel): the core radius grows to accommodate the larger gas mass, thereby smoothing the clump central-peakedness (in other words, $r_c/r_{clump} \simeq (\rho_{edge}/\rho_c)^{1/p_0}$ increases as $m_{clump}$ increases).    Despite the restriction imposed on the initial central density and therefore on the density contrast $\rho_c/\rho_{edge}$, clumps with $p_0=4$ and a mean number density in the range $10^4<n_{H2}<3\cdot10^5\,\cc$ \citep[i.e. the gas densities for which steep density profiles have been detected;][]{schn15} still present magnification factors ranging from $\zeta = 5$ to $25$.  

Based on the results shown in Figs.~\ref{fig:map1}-\ref{fig:rhoclim}, we thus conclude that magnification factors up to at least $\zeta \simeq 25$ are realistic.

\section{Model Consequences}\label{sec:conseq}
\subsection{The measured \sfe per \fft $\effmeas$}
With the magnification factor $\zeta$ given by Eq.~\ref{eq:zeta1}, it is clear that the degree of central concentration of a molecular clump can significantly inflates its globally-measured \sfe per free-fall time $\effmeas$, thereby masking the intrinsic efficiency $\effint$ at work in any clump region small enough to be considered of uniform density.  Even if  all molecular clumps had the same $\effint$, wide variations in the measured efficiency $\effmeas$ are to be expected, reflecting clump structure variations rather than variations of the intrinsic \sfe per free-fall time.  
$\effmeas$ is higher than $\effint$ by the factor $\zeta$, meaning that differences reaching an order of magnitude or more are doable.  In a follow-up paper, we will show how one can estimate the intrinsic \sfe per \fft and, from there, estimate the impact that the initial density profile of a clump has had on its past \stf rate.     

\subsection{Did the progenitors of old globular clusters have a steep density profile?}
\citet{rah19} have studied the ability of star clusters formed at the center of spherically-symmetric molecular clouds to expel the cloud gas as a function of the cloud density profile.  Although the initial steepness $p_0$ of their models (their $-\alpha$ parameter) is not as extreme as in this paper (they range from $0$ to $2$), it is interesting to note that they find that steeper density profiles are less able to unbind the cloud gas.  That is, the outwardly-propagating feedback-driven shell in which the cloud gas is collected is more likely to fall back onto the cluster when the density profile is steep (in their case, when $p_0=2$; see their fig.~4).  If this gas collapse contributes a new episode of star formation in the cluster, and if this scenario repeats itself several times, it could contribute to explaining multiple stellar populations in star clusters \citep{rah18}. 
Steep density profiles inside gaseous cluster progenitors therefore present an interesting two-fold potential:
(i) they favour a high \stf efficiency (see Fig.~\ref{fig:denprotim}) and, thus, the formation of bound clusters; (ii) they may also favour the formation of subsequent/multiple stellar populations, as those now routinely observed in Galactic old globular clusters.       

\section{Conclusions}\label{sec:conc}

Molecular clumps have a higher \sfr when they present a volume density gradient than when they are uniform in density \citep{tan06,gir11a,cho11,elm11,par14p}.  This effect arises from the clump inner regions being denser  than the clump as a whole, yielding faster and more efficient \stf than would be expected based on the clump mean free-fall time (see Figs~\ref{fig:denprotim} and \ref{fig:sfh_2p}).  In this contribution, we have expanded the model of \citet{par13} to map this effect in a systematic and comprehensive way.  For this purpose, we adopt a power-law of slope $-p_0$ and central core radius $r_c$ to describe the initial gas volume-density profile of molecular clumps (Eq.~\ref{eq:denpro}).  We refer to $p_0$ as the steepness of the clump density profile.  We have computed the \sfr of clumps $SFR_{clump}$ (Eq.~\ref{eq:sfrint}) as a function of their mass $m_{clump}$, radius $r_{clump}$, profile steepness $p_0$, and time $t$ since \stf onset.  We adopt an intrinsic \sfe per free-fall time $\effint$ to quantify the \stf activity of any clump region small enough to be considered of uniform density (see Eq.~\ref{eq:sfrint}).  This one is kept constant through all our simulations: $\effint=0.01$.  Our models encompass 4 values of $p_0$: $p_0=1, 2, 3, 4$, the steepest profile being observed in the high-density regions of the nearby molecular cloud NGC6334 \citep{schn15}.  We have run two types of model.  Either we impose the central core radius ($r_c=0.02$\,pc) and we infer the gas central density $\rho_c$ such that a mass $m_{clump}$ is enclosed inside the radius $r_{clump}$ given the adopted initial density profile.  Or we impose the central density ($\rho_c=7\cdot10^6\,\Msppp$) and infer the central core radius $r_c$.    

We refer to the ratio between the \sfr of a centrally-concentrated clump $SFR_{clump}$ and the \sfr $SFR_{TH}$ of its top-hat equivalent as the {\it magnification factor} $\zeta$ (Eq.~\ref{eq:sfrzeta1}).  Gas clumps which are more centrally-concentrated -- either through a smaller relative extent of the central core $r_c/r_{clump}$ or/and through a steeper density profile  -- have a higher magnification factor $\zeta$ at \stf onset (see Fig.~\ref{fig:map1} and the top panels of Fig.~\ref{fig:rhoclim}).  That is, their density gradient allows them to enhance their \sfr with respect to what they would experience should they be of uniform density.  As time goes by, the density profile loses part of its central peakedness (see Fig.~\ref{fig:denprotim}) and the \sfr decreases as a function of time, the decrease being faster for steeper density profiles  (Fig.~\ref{fig:sfh_2p}).  As a result, the magnification factor $\zeta$ decreases with time (compare Figs~\ref{fig:map1} and \ref{fig:map2}, and the top and bottom panels of Fig.~\ref{fig:rhoclim}).                  
 
That a steep density profile amplifies the \sfr of clumps also impacts the estimate of their \sfe per free-fall time.  To infer it, observers often build on the global properties of the gas reservoir under scrutiny, i.e. its mass, mean density (hence free-fall time) and \stf rate.  We refer to this estimate as the {\it measured} \sfe per free-fall time $\effmeas$ (Eq.~\ref{eq:effmeas}).  Global properties, however, do not account for the clump density profile and, therefore, renders an inflated \sfe per \fft compared to what would be found for a uniform-density profile.  
The magnification factor $\zeta = SFR_{clump}/SFR_{TH}$ that we have mapped also equates the ratio between the (globally-)measured \sfe per \fft and its {\it intrinsic} counterpart, i.e. $\zeta=\effmeas/\effint$ (Eq.~\ref{eq:zeta1}).  More centrally-concentrated clumps yield therefore higher measured \stf efficiencies per free-fall time.  
The implications are that, even for a fixed $\effint$, its measured counterpart  $\effmeas$ present wide fluctuations, reflecting the diversity of clump inner structures rather than variations in the process of \stf itself.  Top-hat or shallow profiles have no or little impact on the \sfr and, for them only, the measured \sfe per \fft provides a sensible estimate of the intrinsic efficiency.  
   
With this contribution we therefore encourage observers to become more cognisant of the fact that variations in their measured \sfe per \fft are -- at least partly -- driven by differences in the structure of molecular clumps hosted by the  \sfing regions they survey, rather than by variations in the intrinsic \sfe per free-fall time itself.  Our result for molecular clumps is akin to that found by \citet{lad10} and \citet{lad13} for giant molecular clouds of the Solar neighborhood, as they show that molecular clouds with a larger dense-gas fraction, e.g. more clumps, have a higher \stf rate.  In this contribution we have shown that the gas distribution on the smaller spatial scale of molecular clumps matters as well.    



\acknowledgments
GP is grateful to Douglas Heggie and Anna Pasquali for stimulating discussions while working on this manuscript.  GP also thanks Bruce Elmegreen for having, at the Aspen meeting {\it Modes of Star Formation: a Symposium in Honor of Jay Gallagher}, drawn her attention to the work of \citet{schn15}.  
GP acknowledges support from the Sonderforschungsbereich SFB 881 "The Milky Way System" (subproject B5) of the German Research Foundation (DFG).








\begin{thebibliography}{}

\bibitem[Cho \& Kim (2011)]{cho11}
Cho, W., \& Kim, J. 2011, \mnras, 410, L8

\bibitem[Denissenkov \& Hartwick(2014)]{den14}
Denissenkov, P.A., Hartwick, F. D. A. 2014, \mnras,  437, 21

\bibitem[Elmegreen(2011)]{elm11}
Elmegreen, B.G. 2011, \apj, 731, 61

\bibitem[Elmegreen(2018)]{elm18}
Elmegreen, B.G. 2018, \apj, 854, 16

\bibitem[Evans et al.(2009)]{eva09}
Evans, J.E. II, Dunham, M.M., Jorgensen, J.K., Enoch, M.L., Merin, B. 2009, \apjs, 181, 321

\bibitem[Gao \& Solomon(2004)]{gs04}
Gao, Y; Solomon, P.M. 2004, \apj, 606, 271

\bibitem[Ginsburg et al.(2018)]{gin18}
Ginsburg, A.; Bally, J.; Barnes, A.; Bastian, N.; Battersby, C. 2018, \apj, 853, 171

\bibitem[Girichidis et al.(2011)]{gir11a}
Girichidis, P., Federrath, C., Banerjee, R., Klessen, R.S. 2011, \mnras, 413, pp2741-2759 

\bibitem[Gutermuth et al.(2011)]{gut11}
Gutermuth, R.A., Pipher, J.L., Megeath, S.T., et al. 2011, \apj, 739, 84

\bibitem[Kainulainen et al.~(2014)]{kai14}
Kainulainen, J., Federrath, C., Henning, Th. 2014, Science, 344, 182

\bibitem[{Kritsuk}, {Norman}, \& {Wagner}(2011)]{kri11}
{Kritsuk}, A.~G. and {Norman}, M.~L. and {Wagner}, R. 2011, \apjl, 727, L20

\bibitem[Krumholz \& McKee(2005)]{kru05}
Krumholz, M.R., McKee, C.F. 2005, \apj, 630, 250

\bibitem[Krumholz \& Tan(2007)]{kt07}
Krumholz, M.R., Tan, J.C. 2007, \apj, 654, 304

\bibitem[Krumholz, Dekel \& McKee(2012)]{kru12}
Krumholz, M.R., Dekel, A. \& McKee, C.F. 2012, \apj, 745, 69

\bibitem[Lada, Lombardi \& Alves(2010)]{lad10}
Lada, C.J., Lombardi, M., Alves, J.F. 2010, \apj, 724, 687

\bibitem[Lada et al.(2013)]{lad13}
Lada, C.J., Lombardi, M., Roman-Zuniga, C., Forbrich, J., Alves, J.F. 2013, \apj, 778, 133

\bibitem[Motte et al.(1998)]{mot98}
Motte, F., Andr\'e, P., \& Neri, R. 1998, A\&A, 336, 150

\bibitem[Motte et al.(2018)]{mot18}
Motte, F., Bontemps, S., \& Louvet, F. 2018, ARA\&A, 56, 41

\bibitem[{{M\"uller} {et al.}(2002) {M\"uller}, {Shirley}, {Evans}, {Jacobson} }]{mue02}
{M\"uller}, K.E., {Shirley}, Y.L., {Evans}, N.J. II \& {Jacobson}, H.R. 2002, ApJS, 143, 469 

\bibitem[Murray(2011)]{mur11}
Murray, N. 2011, \apj, 729, 133

\bibitem[Ochsendorf et al.(2017)]{och17}
Ochsendorf, B.~B., Meixner, M., Roman-Duval, J., Rahman, M. \& Evans, II, N.J. 2017, \apj, 841, 109

\bibitem[Parmentier \& Pfalzner(2013)]{par13}
Parmentier, G., Pfalzner, S. 2013, A\&A, 549, 132

\bibitem[Parmentier(2014)]{par14p}
Parmentier, G. 2014, Astronomische Nachrichten, 335, 543

\bibitem[Parmentier, Pfalzner \& Grebel(2014)]{par14}
Parmentier, G., Pfalzner, S. \& Grebel E.K. 2014, \apj, 791, 132

\bibitem[Parmentier(2016)]{par16}
Parmentier, G. 2016, \apj, 826, 74

\bibitem[Parmentier(2017)]{par17}
Parmentier, G. 2017, \apj, 843, 7

\bibitem[Rahner et al.(2018)]{rah18}
Rahner, D., Pellegrini, E.W., Glover, S.C.O., Klessen, R.S. 2018, \mnras, 473, L11-L15

\bibitem[Rahner et al.(2019)]{rah19}
Rahner, D., Pellegrini, E.W., Glover, S.C.O., Klessen, R.S. 2019, \mnras, 483, pp2547-2560

\bibitem[Schneider et al.(2015)]{schn15}
Schneider, N., Bontemps, S., Girichidis, P., Rayner, T., Motte, F., et al 2015, \mnras, 453, L41-L45

\bibitem[V\'azquez-Semadeni(1994)]{vs94}
V\'azquez-Semadeni, E. 1994, \apj, 423, 681

\bibitem[Tan et al.(2006)]{tan06}
Tan, J.C., Krumholz, M.R. \& McKee, C.F. 2006, \apjl, 641, L121

\bibitem[Vutisalchavakul, Evans \& Heyer(2016)]{vut16}
Vutisalchavakul, N.; Evans, N.J., II \& Heyer, M. 2016, \apj, 831, 73

\bibitem[Watkins et al.(2019)]{wat19}
Watkins, E.J.; Peretto, N.; Marsch, K.; Fuller, G.A. 2019 A\&A 628, 21

\bibitem[Wu et al.(2010)]{wu10}
Wu, J., Evans, N.J., II, Shirley, Y.L., Knez, C. 2010, ApJS, 188, 313 

\end{thebibliography}
\end{document}